\documentclass[11pt, nofonttune]{IEEEtran}
\usepackage{url}
\usepackage[utf8]{inputenc}
\usepackage{xcolor}
\usepackage{amsmath}
\usepackage{amssymb}
\usepackage{xspace}
\usepackage{epsfig}
\usepackage{balance}
\usepackage{placeins}

\usepackage[acronyms,nonumberlist,nopostdot,nomain,nogroupskip,acronymlists={hidden}]{glossaries}
\newglossary[algh]{hidden}{acrh}{acnh}{Hidden Acronyms}
\usepackage{tablefootnote}
\usepackage{booktabs}

\usepackage{tabularx}
\usepackage{hyperref}
\usepackage{soul}

\makeatletter
\let\oldmakefirstuc\makefirstuc
\renewcommand*{\makefirstuc}[1]{%
  \def\gls@add@space{}%
  \mfu@capitalisewords#1 \@nil\mfu@endcap
}
\def\mfu@capitalisewords#1 #2\mfu@endcap{%
  \def\mfu@cap@first{#1}%
  \def\mfu@cap@second{#2}%
  \gls@add@space
  \oldmakefirstuc{#1}%
  \def\gls@add@space{ }%
  \ifx\mfu@cap@second\@nnil
    \let\next@mfu@cap\mfu@noop
  \else
    \let\next@mfu@cap\mfu@capitalisewords
  \fi
  \next@mfu@cap#2\mfu@endcap
}
\makeatother

\usepackage{multirow}

\usepackage[font=footnotesize]{subcaption}
\usepackage[font=footnotesize]{caption}

\usepackage{mathtools}
\usepackage{cite}

\newacronym{3gpp}{3GPP}{3rd Generation Partnership Project}
\newacronym{4g}{4G}{4th generation}
\newacronym{5g}{5G}{5th generation}
\newacronym{5gc}{5GC}{5G Core}
\newacronym{adc}{ADC}{Analog to Digital Converter}
\newacronym{aerpaw}{AERPAW}{Aerial Experimentation and Research Platform for Advanced Wireless}
\newacronym{ai}{AI}{Artificial Intelligence}
\newacronym{aimd}{AIMD}{Additive Increase Multiplicative Decrease}
\newacronym{am}{AM}{Acknowledged Mode}
\newacronym{amc}{AMC}{Adaptive Modulation and Coding}
\newacronym{amf}{AMF}{Access and Mobility Management Function}
\newacronym{aops}{AOPS}{Adaptive Order Prediction Scheduling}
\newacronym{api}{API}{Application Programming Interface}
\newacronym{apn}{APN}{Access Point Name}
\newacronym{ap}{AP}{application protocol}
\newacronym{aqm}{AQM}{Active Queue Management}
\newacronym{ausf}{AUSF}{Authentication Server Function}
\newacronym{avc}{AVC}{Advanced Video Coding}
\newacronym{awgn}{AGWN}{Additive White Gaussian Noise}
\newacronym{balia}{BALIA}{Balanced Link Adaptation Algorithm}
\newacronym{bbu}{BBU}{Base Band Unit}
\newacronym{bdp}{BDP}{Bandwidth-Delay Product}
\newacronym{ber}{BER}{Bit Error Rate}
\newacronym{bf}{BF}{Beamforming}
\newacronym{bler}{BLER}{Block Error Rate}
\newacronym{brr}{BRR}{Bayesian Ridge Regressor}
\newacronym{bs}{BS}{Base Station}
\newacronym{bsr}{BSR}{Buffer Status Report}
\newacronym{bss}{BSS}{Business Support System}
\newacronym{ca}{CA}{Carrier Aggregation}
\newacronym{caas}{CaaS}{Connectivity-as-a-Service}
\newacronym{cb}{CB}{Code Block}
\newacronym{cc}{CC}{Congestion Control}
\newacronym{ccid}{CCID}{Congestion Control ID}
\newacronym{cco}{CC}{Carrier Component}
\newacronym{cdd}{CDD}{Cyclic Delay Diversity}
\newacronym{cdf}{CDF}{Cumulative Distribution Function}
\newacronym{cdn}{CDN}{Content Distribution Network}
\newacronym{cn}{CN}{Core Network}
\newacronym{codel}{CoDel}{Controlled Delay Management}
\newacronym{comac}{COMAC}{Converged Multi-Access and Core}
\newacronym{cord}{CORD}{Central Office Re-architected as a Datacenter}
\newacronym{cornet}{CORNET}{COgnitive Radio NETwork}
\newacronym{cosmos}{COSMOS}{Cloud Enhanced Open Software Defined Mobile Wireless Testbed for City-Scale Deployment}
\newacronym{cots}{COTS}{Commercial Off-the-Shelf}
\newacronym{cp}{CP}{Control Plane}
\newacronym{cpu}{CPU}{Central Processing Unit}
\newacronym{cqi}{CQI}{Channel Quality Information}
\newacronym{cr}{CR}{Cognitive Radio}
\newacronym{cran}{CRAN}{Cloud \gls{ran}}
\newacronym{crs}{CRS}{Cell Reference Signal}
\newacronym{csi}{CSI}{Channel State Information}
\newacronym{csirs}{CSI-RS}{Channel State Information - Reference Signal}
\newacronym{cu}{CU}{Central Unit}
\newacronym{d2tcp}{D$^2$TCP}{Deadline-aware Data center TCP}
\newacronym{d3}{D$^3$}{Deadline-Driven Delivery}
\newacronym{dac}{DAC}{Digital to Analog Converter}
\newacronym{dag}{DAG}{Directed Acyclic Graph}
\newacronym{das}{DAS}{Distributed Antenna System}
\newacronym{dash}{DASH}{Dynamic Adaptive Streaming over HTTP}
\newacronym{dc}{DC}{Dual Connectivity}
\newacronym{dccp}{DCCP}{Datagram Congestion Control Protocol}
\newacronym{dce}{DCE}{Direct Code Execution}
\newacronym{dci}{DCI}{Downlink Control Information}
\newacronym{dctcp}{DCTCP}{Data Center TCP}
\newacronym{dl}{DL}{Downlink}
\newacronym{dmr}{DMR}{Deadline Miss Ratio}
\newacronym{dmrs}{DMRS}{DeModulation Reference Signal}
\newacronym{drlcc}{DRL-CC}{Deep Reinforcement Learning Congestion Control}
\newacronym{drs}{DRS}{Discovery Reference Signal}
\newacronym{du}{DU}{Distributed Unit}
\newacronym{e2e}{E2E}{end-to-end}
\newacronym{ecaas}{ECaaS}{Edge-Cloud-as-a-Service}
\newacronym{ecn}{ECN}{Explicit Congestion Notification}
\newacronym{edf}{EDF}{Earliest Deadline First}
\newacronym{embb}{eMBB}{Enhanced Mobile Broadband}
\newacronym{empower}{EMPOWER}{EMpowering transatlantic PlatfOrms for advanced WirEless Research}
\newacronym{enb}{eNB}{evolved Node Base}
\newacronym{endc}{EN-DC}{E-UTRAN-\gls{nr} \gls{dc}}
\newacronym{epc}{EPC}{Evolved Packet Core}
\newacronym{eps}{EPS}{Evolved Packet System}
\newacronym{es}{ES}{Edge Server}
\newacronym{etsi}{ETSI}{European Telecommunications Standards Institute}
\newacronym[firstplural=Estimated Times of Arrival (ETAs)]{eta}{ETA}{Estimated Time of Arrival}
\newacronym{eutran}{E-UTRAN}{Evolved Universal Terrestrial Access Network}
\newacronym{faas}{FaaS}{Function-as-a-Service}
\newacronym{fapi}{FAPI}{Functional Application Platform Interface}
\newacronym{fdd}{FDD}{Frequency Division Duplexing}
\newacronym{fdm}{FDM}{Frequency Division Multiplexing}
\newacronym{fdma}{FDMA}{Frequency Division Multiple Access}
\newacronym{fed4fire}{FED4FIRE+}{Federation 4 Future Internet Research and Experimentation Plus}
\newacronym{fir}{FIR}{finite impulse response}
\newacronym{fit}{FIT}{Future \acrlong{iot}}
\newacronym{fpga}{FPGA}{Field Programmable Gate Array}
\newacronym{fr2}{FR2}{Frequency Range 2}
\newacronym{fs}{FS}{Fast Switching}
\newacronym{fscc}{FSCC}{Flow Sharing Congestion Control}
\newacronym{ftp}{FTP}{File Transfer Protocol}
\newacronym{fw}{FW}{Flow Window}
\newacronym{ge}{GE}{Gaussian Elimination}
\newacronym{gnb}{gNB}{Next Generation Node Base}
\newacronym{gop}{GOP}{Group of Pictures}
\newacronym{gpr}{GPR}{Gaussian Process Regressor}
\newacronym{gpu}{GPU}{Graphics Processing Unit}
\newacronym{gtp}{GTP}{GPRS Tunneling Protocol}
\newacronym{gtpc}{GTP-C}{GPRS Tunnelling Protocol Control Plane}
\newacronym{gtpu}{GTP-U}{GPRS Tunnelling Protocol User Plane}
\newacronym{gtpv2c}{GTPv2-C}{\gls{gtp} v2 - Control}
\newacronym{gw}{GW}{Gateway}
\newacronym{harq}{HARQ}{Hybrid Automatic Repeat reQuest}
\newacronym{hetnet}{HetNet}{Heterogeneous Network}
\newacronym{hh}{HH}{Hard Handover}
\newacronym{hol}{HOL}{Head-of-Line}
\newacronym{hqf}{HQF}{Highest-quality-first}
\newacronym{hss}{HSS}{Home Subscription Server}
\newacronym{http}{HTTP}{HyperText Transfer Protocol}
\newacronym{ia}{IA}{Initial Access}
\newacronym{iab}{IAB}{Integrated Access and Backhaul}
\newacronym{ic}{IC}{Incident Command}
\newacronym{ietf}{IETF}{Internet Engineering Task Force}
\newacronym{imsi}{IMSI}{International Mobile Subscriber Identity}
\newacronym{imt}{IMT}{International Mobile Telecommunication}
\newacronym{iot}{IoT}{Internet of Things}
\newacronym{ip}{IP}{Internet Protocol}
\newacronym{itu}{ITU}{International Telecommunication Union}
\newacronym{kpi}{KPI}{Key Performance Indicator}
\newacronym{kpm}{KPM}{Key Performance Measurement}
\newacronym{kvm}{KVM}{Kernel-based Virtual Machine}
\newacronym{los}{LOS}{Line-of-Sight}
\newacronym{lsm}{LSM}{Link-to-System Mapping}
\newacronym{lstm}{LSTM}{Long Short Term Memory}
\newacronym{lte}{LTE}{Long Term Evolution}
\newacronym{lxc}{LXC}{Linux Containers}
\newacronym{m2m}{M2M}{Machine to Machine}
\newacronym{mac}{MAC}{Medium Access Control}
\newacronym{manet}{MANET}{Mobile Ad Hoc Network}
\newacronym{mano}{MANO}{management~and orchestration}
\newacronym{mc}{MC}{Multi-Connectivity}
\newacronym{mcc}{MCC}{Mobile Cloud Computing}
\newacronym{mchem}{MCHEM}{Massive Channel Emulator}
\newacronym{mcs}{MCS}{Modulation and Coding Scheme}
\newacronym{mec}{MEC}{Multi-access Edge Computing}
\newacronym{mec2}{MEC}{Mobile Edge Cloud}
\newacronym{mfc}{MFC}{Mobile Fog Computing}
\newacronym{mi}{MI}{Mutual Information}
\newacronym{mib}{MIB}{Master Information Block}
\newacronym{miesm}{MIESM}{Mutual Information Based Effective SINR}
\newacronym{mimo}{MIMO}{Multiple Input, Multiple Output}
\newacronym{ml}{ML}{Machine Learning}
\newacronym{mlr}{MLR}{Maximum-local-rate}
\newacronym[plural=\gls{mme}s,firstplural=Mobility Management Entities (MMEs)]{mme}{MME}{Mobility Management Entity}
\newacronym{mmtc}{mMTC}{Massive Machine-Type Communications}
\newacronym{mmwave}{mmWave}{millimeter wave}
\newacronym{mpdccp}{MP-DCCP}{Multipath Datagram Congestion Control Protocol}
\newacronym{mptcp}{MPTCP}{Multipath TCP}
\newacronym{mr}{MR}{Maximum Rate}
\newacronym{mrdc}{MR-DC}{Multi \gls{rat} \gls{dc}}
\newacronym{mse}{MSE}{Mean Square Error}
\newacronym{mss}{MSS}{Maximum Segment Size}
\newacronym{mt}{MT}{Mobile Termination}
\newacronym{mtd}{MTD}{Machine-Type Device}
\newacronym{mtu}{MTU}{Maximum Transmission Unit}
\newacronym{mumimo}{MU-MIMO}{Multi-user \gls{mimo}}
\newacronym{mvno}{MVNO}{Mobile Virtual Network Operator}
\newacronym{nalu}{NALU}{Network Abstraction Layer Unit}
\newacronym{nas}{NAS}{Non-Access Stratum}
\newacronym{nbiot}{NB-IoT}{Narrow Band IoT}
\newacronym{nfv}{NFV}{network function virtualization}
\newacronym{nfvi}{NFVI}{Network Function Virtualization Infrastructure}
\newacronym{nic}{NIC}{Network Interface Card}
\newacronym{nlos}{NLOS}{Non-Line-of-Sight}
\newacronym{now}{NOW}{Non Overlapping Window}
\newacronym{nsm}{NSM}{Network Service Mesh}
\newacronym[type=hidden]{nr}{NR}{New Radio}
\newacronym{nrf}{NRF}{Network Repository Function}
\newacronym{nsa}{NSA}{Non Stand Alone}
\newacronym{nse}{NSE}{Network Slicing Engine}
\newacronym{nssf}{NSSF}{Network Slice Selection Function}
\newacronym{o2i}{O2I}{Outdoor to Indoor}
\newacronym{oai}{OAI}{OpenAirInterface}
\newacronym{oaicn}{OAI-CN}{\gls{oai} \acrlong{cn}}
\newacronym{oairan}{OAI-RAN}{\acrlong{oai} \acrlong{ran}}
\newacronym{oam}{OAM}{Operations, Administration and Maintenance}
\newacronym{ofdm}{OFDM}{Orthogonal Frequency Division Multiplexing}
\newacronym{olia}{OLIA}{Opportunistic Linked Increase Algorithm}
\newacronym{omec}{OMEC}{Open Mobile Evolved Core}
\newacronym{onap}{ONAP}{Open Network Automation Platform}
\newacronym{onf}{ONF}{Open Networking Foundation}
\newacronym{onos}{ONOS}{Open Networking Operating System}
\newacronym{oom}{OOM}{\gls{onap} Operations Manager}
\newacronym{opnfv}{OPNFV}{Open Platform for \gls{nfv}}
\newacronym[type=hidden]{oran}{O-RAN}{Open \gls{ran}}
\newacronym{orbit}{ORBIT}{Open-Access Research Testbed for Next-Generation Wireless Networks}
\newacronym{os}{OS}{Operating System}
\newacronym{oss}{OSS}{Operations Support System}
\newacronym{pa}{PA}{Position-aware}
\newacronym{pase}{PASE}{Prioritization, Arbitration, and Self-adjusting Endpoints}
\newacronym{pawr}{PAWR}{Platforms for Advanced Wireless Research}
\newacronym{pbch}{PBCH}{Physical Broadcast Channel}
\newacronym{pcef}{PCEF}{Policy and Charging Enforcement Function}
\newacronym{pcfich}{PCFICH}{Physical Control Format Indicator Channel}
\newacronym{pcrf}{PCRF}{Policy and Charging Rules Function}
\newacronym{pdcch}{PDCCH}{Physical Downlink Control Channel}
\newacronym{pdcp}{PDCP}{Packet Data Convergence Protocol}
\newacronym{pdsch}{PDSCH}{Physical Downlink Shared Channel}
\newacronym{pdu}{PDU}{Packet Data Unit}
\newacronym{pf}{PF}{Proportional Fair}
\newacronym{pgw}{PGW}{Packet Gateway}
\newacronym{phich}{PHICH}{Physical Hybrid ARQ Indicator Channel}
\newacronym{phy}{PHY}{Physical}
\newacronym{pmch}{PMCH}{Physical Multicast Channel}
\newacronym{pmi}{PMI}{Precoding Matrix Indicators}
\newacronym{powder}{POWDER}{Platform for Open Wireless Data-driven Experimental Research}
\newacronym{ppo}{PPO}{Proximal Policy Optimization}
\newacronym{ppp}{PPP}{Poisson Point Process}
\newacronym{prach}{PRACH}{Physical Random Access Channel}
\newacronym{prb}{PRB}{Physical Resource Block}
\newacronym{psnr}{PSNR}{Peak Signal to Noise Ratio}
\newacronym{pss}{PSS}{Primary Synchronization Signal}
\newacronym{pucch}{PUCCH}{Physical Uplink Control Channel}
\newacronym{pusch}{PUSCH}{Physical Uplink Shared Channel}
\newacronym{qam}{QAM}{Quadrature Amplitude Modulation}
\newacronym{qci}{QCI}{\gls{qos} Class Identifier}
\newacronym{qoe}{QoE}{Quality of Experience}
\newacronym{qos}{QoS}{Quality of Service}
\newacronym{quic}{QUIC}{Quick UDP Internet Connections}
\newacronym{rach}{RACH}{Random Access Channel}
\newacronym{ran}{RAN}{Radio Access Network}
\newacronym[firstplural=Radio Access Technologies (RATs)]{rat}{RAT}{Radio Access Technology}
\newacronym{rcn}{RCN}{Research Coordination Network}
\newacronym{rec}{REC}{Radio Edge Cloud}
\newacronym{red}{RED}{Random Early Detection}
\newacronym{renew}{RENEW}{Reconfigurable Eco-system for Next-generation End-to-end Wireless}
\newacronym{rf}{RF}{Radio Frequency}
\newacronym{rfc}{RFC}{Request for Comments}
\newacronym{rfr}{RFR}{Random Forest Regressor}
\newacronym{ric}{RIC}{\gls{ran} Intelligent Controller}
\newacronym{rlc}{RLC}{Radio Link Control}
\newacronym{rlf}{RLF}{Radio Link Failure}
\newacronym{rlnc}{RLNC}{Random Linear Network Coding}
\newacronym{rmr}{RMR}{RIC Message Router}
\newacronym{rmse}{RMSE}{Root Mean Squared Error}
\newacronym{rnis}{RNIS}{Radio Network Information Service}
\newacronym{rr}{RR}{Round Robin}
\newacronym{rrc}{RRC}{Radio Resource Control}
\newacronym{rrm}{RRM}{Radio Resource Management}
\newacronym{rru}{RRU}{Remote Radio Unit}
\newacronym{rs}{RS}{Remote Server}
\newacronym{rsrp}{RSRP}{Reference Signal Received Power}
\newacronym{rsrq}{RSRQ}{Reference Signal Received Quality}
\newacronym{rss}{RSS}{Received Signal Strength}
\newacronym{rssi}{RSSI}{Received Signal Strength Indicator}
\newacronym{rtt}{RTT}{Round Trip Time}
\newacronym{ru}{RU}{Radio Unit}
\newacronym{rw}{RW}{Receive Window}
\newacronym{rx}{RX}{Receiver}
\newacronym{s1ap}{S1AP}{S1 Application Protocol}
\newacronym{sa}{SA}{standalone}
\newacronym{sack}{SACK}{Selective Acknowledgment}
\newacronym{sap}{SAP}{Service Access Point}
\newacronym{sc2}{SC2}{Spectrum Collaboration Challenge}
\newacronym{scef}{SCEF}{Service Capability Exposure Function}
\newacronym{sch}{SCH}{Secondary Cell Handover}
\newacronym{scoot}{SCOOT}{Split Cycle Offset Optimization Technique}
\newacronym{sctp}{SCTP}{Stream Control Transmission Protocol}
\newacronym{sdap}{SDAP}{Service Data Adaptation Protocol}
\newacronym{sdk}{SDK}{Software Development Kit}
\newacronym{sdm}{SDM}{Space Division Multiplexing}
\newacronym{sdma}{SDMA}{Spatial Division Multiple Access}
\newacronym{sdn}{SDN}{Software-defined Networking}
\newacronym{sdr}{SDR}{Software-defined Radio}
\newacronym{seba}{SEBA}{SDN-Enabled Broadband Access}
\newacronym{sgsn}{SGSN}{Serving GPRS Support Node}
\newacronym{sgw}{SGW}{Service Gateway}
\newacronym{si}{SI}{Study Item}
\newacronym{sib}{SIB}{Secondary Information Block}
\newacronym{sinr}{SINR}{Signal to Interference plus Noise Ratio}
\newacronym{sip}{SIP}{Session Initiation Protocol}
\newacronym{siso}{SISO}{Single Input, Single Output}
\newacronym{sla}{SLA}{service level agreement}
\newacronym{sm}{SM}{Service Model}
\newacronym{smf}{SMF}{Session Management Function}
\newacronym{smo}{SMO}{service management and orchestration}
\newacronym{sms}{SMS}{Short Message Service}
\newacronym{smsgmsc}{SMS-GMSC}{\gls{sms}-Gateway}
\newacronym{snr}{SNR}{Signal-to-Noise-Ratio}
\newacronym{son}{SON}{Self-Organizing Network}
\newacronym{sptcp}{SPTCP}{Single Path TCP}
\newacronym{srb}{SRB}{Service Radio Bearer}
\newacronym{srn}{SRN}{Standard Radio Node}
\newacronym{srs}{SRS}{Sounding Reference Signal}
\newacronym{ss}{SS}{Synchronization Signal}
\newacronym{sss}{SSS}{Secondary Synchronization Signal}
\newacronym{st}{ST}{Spanning Tree}
\newacronym{svc}{SVC}{Scalable Video Coding}
\newacronym{tb}{TB}{Transport Block}
\newacronym{tcp}{TCP}{Transmission Control Protocol}
\newacronym{tdd}{TDD}{Time Division Duplexing}
\newacronym{tdm}{TDM}{Time Division Multiplexing}
\newacronym{tdma}{TDMA}{Time Division Multiple Access}
\newacronym{tfl}{TfL}{Transport for London}
\newacronym{tfrc}{TFRC}{TCP-Friendly Rate Control}
\newacronym{tft}{TFT}{Traffic Flow Template}
\newacronym{tgen}{TGEN}{Traffic Generator}
\newacronym{tip}{TIP}{Telecom Infra Project}
\newacronym{tm}{TM}{Transparent Mode}
\newacronym{to}{TO}{Telco Operator}
\newacronym{tr}{TR}{Technical Report}
\newacronym{trp}{TRP}{Transmitter Receiver Pair}
\newacronym{ts}{TS}{Technical Specification}
\newacronym{tti}{TTI}{Transmission Time Interval}
\newacronym{ttt}{TTT}{Time-to-Trigger}
\newacronym{tx}{TX}{Transmitter}
\newacronym{uas}{UAS}{Unmanned Aerial System}
\newacronym{uav}{UAV}{Unmanned Aerial Vehicle}
\newacronym{udm}{UDM}{Unified Data Management}
\newacronym{udp}{UDP}{User Datagram Protocol}
\newacronym{udr}{UDR}{Unified Data Repository}
\newacronym{ue}{UE}{User Equipment}
\newacronym{uhd}{UHD}{\gls{usrp} Hardware Driver}
\newacronym{ul}{UL}{Uplink}
\newacronym{um}{UM}{Unacknowledged Mode}
\newacronym{uml}{UML}{Unified Modeling Language}
\newacronym{upa}{UPA}{Uniform Planar Array}
\newacronym{upf}{UPF}{User Plane Function}
\newacronym{urllc}{URLLC}{Ultra Reliable and Low Latency Communication}
\newacronym{usa}{U.S.}{United States}
\newacronym{usim}{USIM}{Universal Subscriber Identity Module}
\newacronym{usrp}{USRP}{Universal Software Radio Peripheral}
\newacronym{utc}{UTC}{Urban Traffic Control}
\newacronym{vim}{VIM}{Virtualization Infrastructure Manager}
\newacronym{vm}{VM}{Virtual Machine}
\newacronym{vnf}{VNF}{Virtual Network Function}
\newacronym{volte}{VoLTE}{Voice over \gls{lte}}
\newacronym{voltha}{VOLTHA}{Virtual OLT HArdware Abstraction}
\newacronym{vr}{VR}{Virtual Reality}
\newacronym{vran}{vRAN}{Virtualized \gls{ran}}
\newacronym{vss}{VSS}{Video Streaming Server}
\newacronym{wbf}{WBF}{Wired Bias Function}
\newacronym{wf}{WF}{Wired-first}
\newacronym{wlan}{WLAN}{Wireless Local Area Network}
\newacronym{osm}{OSM}{Open Source \gls{nfv} Management and Orchestration}
\newacronym{pnf}{PNF}{Physical Network Function}
\newacronym{drl}{DRL}{Deep Reinforcement Learning}
\newacronym{mtc}{MTC}{Machine-type Communications}

\definecolor{desireRed}{RGB}{230,57,60}%
\definecolor{darkPurple}{RGB}{59,31,43}%
\definecolor{springGreen}{RGB}{37,223,145}%
\definecolor{queenBlue}{RGB}{69,123,157}%
\definecolor{spaceCadet}{RGB}{29,53,87}%

\begin{document}
\bstctlcite{BSTcontrol}  % run this directive at start of document

\title{\vspace{-4pt}Intelligence and Learning in O-RAN\\for Data-driven NextG Cellular Networks}

\author{\IEEEauthorblockN{Leonardo Bonati, Salvatore D'Oro, Michele Polese, Stefano Basagni, Tommaso Melodia}
\thanks{The authors are with the Institute for the Wireless Internet of Things, Northeastern University, Boston, MA, USA. E-mail: \{l.bonati, s.doro, m.polese, s.basagni, t.melodia\}@northeastern.edu.}
\thanks{This work was partially supported by the U.S.\ National Science Foundation under Grant CNS-1923789 and the U.S.\ Office of Naval Research under Grant N00014-20-1-2132.}
}

% decrease space after author block
\makeatletter
\patchcmd{\@maketitle}
  {\addvspace{0.5\baselineskip}\egroup}
  {\addvspace{-1.75\baselineskip}\egroup}
  {}
  {}
\makeatother

\flushbottom
\setlength{\parskip}{0ex plus0.1ex}

\maketitle
\glsunset{nr}
\glsunset{lte}

\begin{abstract}
Next Generation (NextG) cellular networks will be natively cloud-based and built upon programmable, virtualized, and disaggregated architectures. The separation of control functions from the hardware fabric and the introduction of standardized control interfaces will enable the definition of custom closed-control loops, which will ultimately enable embedded intelligence and real-time analytics, thus effectively realizing the vision of autonomous and self-optimizing networks.
This article explores the disaggregated network architecture proposed by the \emph{O-RAN Alliance} as a key enabler of NextG networks. Within this architectural context, we discuss the potential, the challenges, and the limitations of data-driven optimization approaches to network control over different timescales. We also present the first large-scale integration of O-RAN-compliant software components with an open-source full-stack softwarized cellular network. Experiments conducted on Colosseum, the world's largest wireless network emulator, demonstrate closed-loop integration of real-time analytics and control through deep reinforcement learning agents. We also show the feasibility of Radio Access Network (RAN) control through xApps running on the near real-time RAN Intelligent Controller, to optimize the scheduling policies of co-existing network slices, leveraging the O-RAN open interfaces to collect data at the edge of the network.
\end{abstract}

% \begin{IEEEkeywords}O-RAN, Network Intelligence, 5G/6G, Machine Learning.
% \end{IEEEkeywords}

\begin{picture}(0,0)(10,-420)
\put(0,0){
\put(0,10){\footnotesize This paper has been accepted for publication on IEEE Communications Magazine.}
\put(0,0){\tiny \copyright 2021 IEEE. Personal use of this material is permitted. Permission from IEEE must be obtained for all other uses, in any current or future media including reprinting/republishing}
\put(0,-6){\tiny this material for advertising or promotional purposes, creating new collective works, for resale or redistribution to servers or lists, or reuse of any copyrighted component of this work in other works.}}
\end{picture}

\glsresetall

\vspace{-0.6cm}
\section{Introduction}
\label{sec:intro}

The fifth (5G) and sixth generations (6G) of cellular networks will undoubtedly accelerate the transition from inflexible and monolithic networks to agile, disaggregated architectures based on \emph{softwarization} and \emph{virtualization}, as well as on openness and re-programmability of network components~\cite{bonati2020open}.
These novel architectures are expected to become enablers of new functionalities, including the ability to: (i)~Provide on-demand virtual network slices that, albeit sharing the same physical infrastructure, are tailored to different mobile virtual network operators, network services and run-time traffic requirements; (ii)~split network functions across multiple software and hardware components, possibly provided by multiple vendors; (iii)~capture and expose \glspl{kpi} and network analytics through open interfaces that are not available in old architectures, and (iv)~control the entire network physical infrastructure in real time via third party software applications and open interfaces.

%\vspace{-0.51cm}
\subsection{Disaggregation and Programmability in O-RAN}

%In this context, 
The \emph{O-RAN Alliance}---a consortium of industry and academic institutions---is working toward realizing the vision of Next Generation (NextG) cellular networks, where telecom operators use standardized interfaces to control multi-vendor infrastructures and deliver high performance services to their subscribers~\cite{oran-arch-spec}.
To achieve this goal, the Alliance proposes an architectural innovation based on two core principles. First, O-RAN embraces and promotes the \gls{3gpp} functional split, where \gls{bs} functionalities are virtualized as network functions and divided across multiple network nodes, i.e., \gls{cu}, \gls{du} and \gls{ru}~\cite{bonati2020open}. 
This facilitates the instantiation and execution of diverse networking processes at different points of the network. Specifically, \glspl{cu} implement functionalities at the higher layers of the protocol stack operating over larger timescales, while \glspl{du} handle time-critical operations at the lower layers. Finally, the \glspl{ru} manage \gls{rf} components and lower \gls{phy} layer parts.

% which is likely to be even more impactful on NextG networks
The second core innovation---which is likely to be even more impactful---is the \gls{ric}, a new architectural component that provides a centralized abstraction of the network, allowing operators to implement and deploy custom control plane functions.
In both its \textit{non} and \textit{near real-time} versions, the \gls{ric} facilitates \gls{ran} optimization through closed-control loops, i.e., autonomous action and feedback loops between \gls{ran} components and their controllers.
O-RAN envisions different
%control
loops operating at timescales that range from $1\:\mathrm{ms}$ (e.g., for real-time control of transmission strategies)
%and beamforming)
to
%hundreds or
thousands of milliseconds (e.g., for network slicing and traffic forecasting).
%and hand-over management).
For instance, the non real-time \gls{ric} performs operations with a time granularity higher than one second, such as training of \gls{ai} and \gls{ml} models.
%and service provisioning.
%
The near real-time \gls{ric} instead handles procedures at timescales above $10\:\mathrm{ms}$, hosts third party applications (\textit{xApps}) that communicate with the \gls{cu}/\gls{du} through standardized open interfaces, and implements intelligence in the \gls{ran} through data-driven control loops.

\begin{figure}[ht]
	\centering
	\includegraphics[width=\columnwidth]{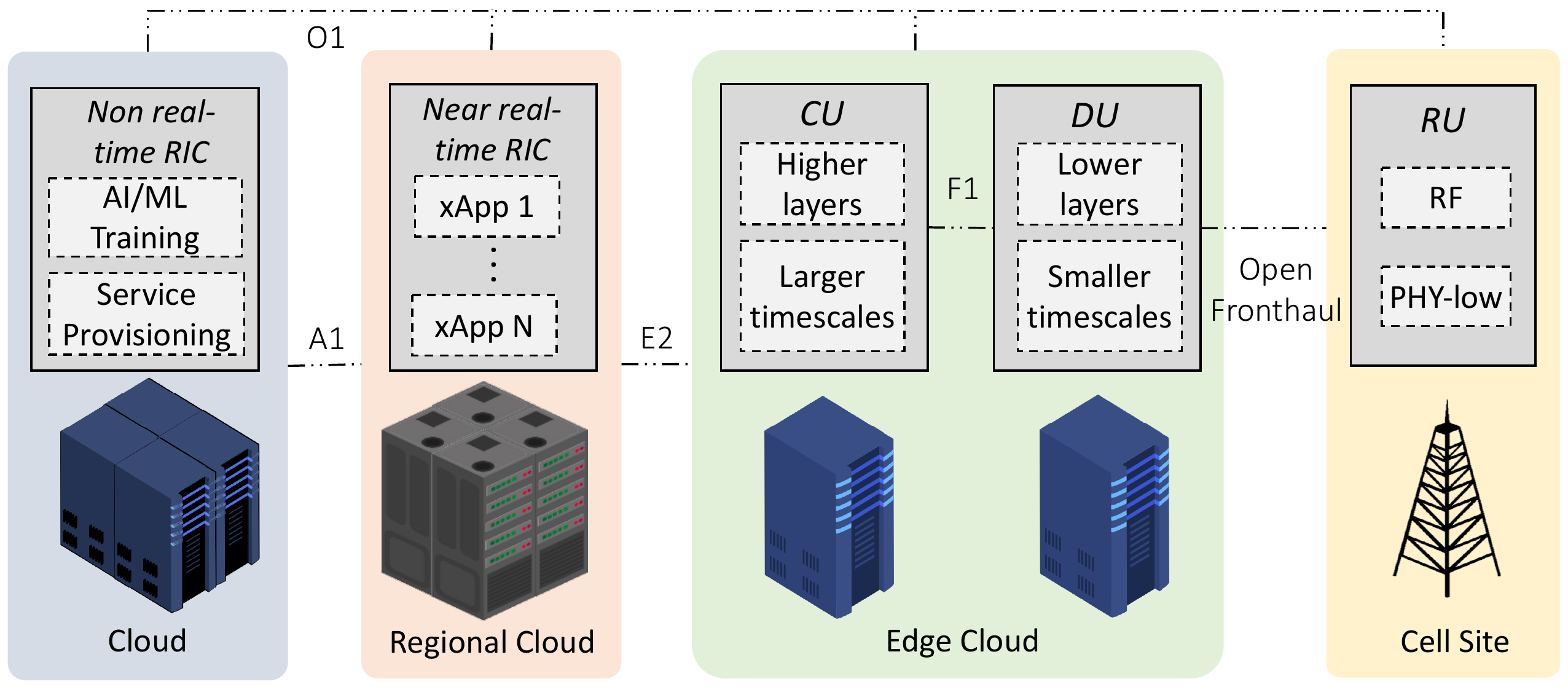}
	\caption{O-RAN: An example of disaggregated deployment.}
	\label{fig:deployment}
\end{figure}

Figure~\ref{fig:deployment} illustrates one of the possible disaggregated deployments specified by O-RAN, where different network components are connected by open interfaces.

In this deployment (``Scenario~B'', deemed the most common~\cite{bonati2020open}), the \glspl{ric} are deployed in the cloud. They interact with each other via the A1 and O1 interfaces, and control specific parameters of the \gls{ran} defined through the so-called \glspl{sm}.
The \gls{cu} and \gls{du} are deployed at the network edge, interconnected through the F1 interface and controlled by the near real-time \gls{ric} via the E2 interface~\cite{oran-e2}. The \gls{ru} is located at the operator cell site, and controlled by the \gls{du} through the Open Fronthaul interface. Finally, \gls{cu}, \gls{du} and \gls{ru} are connected to the non real-time \gls{ric} through the O1 interface for periodic reporting.
Other deployment options allow to instantiate \glspl{ric} and \glspl{cu} in the edge or regional cloud (Scenarios~A and C-F, respectively); the \gls{du} in the edge cloud (A-D); and the \gls{ru} at the operator cell site (A-D) or in the cloud cell site, possibly co-located with the \gls{du} (E, F)~\cite{bonati2020open}.

\vspace{-0.4cm}
\subsection{Contributions}

While the O-RAN architectural vision is gaining momentum among researchers,
%and practitioners,
the challenges of implementing it for data-driven, open, programmable and virtualized NextG networks are still largely to be dealt with. 
Important architectural questions are yet to be answered, including (i)~the exact functionalities and parameters to be controlled by each network component; (ii)~where to place network intelligence; (iii)~how to validate and train data-driven control loop solutions, and (iv)~how \gls{ai} agents can access data and analytics from the \gls{ran} while minimizing the overhead of moving them from the \gls{ran} to the storage and inference locations.
To answer these questions, we provide the following contributions.
%Aiming at answering these questions, this article provides the following contributions.

\noindent
$\bullet$ We discuss how data-driven, closed-control loop solutions can be implemented in NextG \glspl{ran}. We focus on the \textit{opportunities} offered by the O-RAN architecture, including functional split and open interfaces, and on their role in advancing intelligent and programmable
%wireless
networks.

\noindent
$\bullet$ 
Differently from prior work~\cite{niknam2020intelligent, bonati2020open},
we investigate the \textit{limitations} of the current O-RAN specifications and the \textit{challenges} associated with deploying data-driven policies at different nodes of the \gls{ran}.

\noindent
$\bullet$ We discuss how large-scale experimental testbeds will play a key role by providing researchers with heterogeneous and large datasets, critical to the success of data-driven solutions for cellular networks.
%In particular,
We focus on the three PAWR platforms, i.e., POWDER~\cite{breen2020powder}, COSMOS~\cite{raychaudhuri2020cosmos}, and AERPAW~\cite{sichitiu2020aerpaw}, and on Colosseum and Arena~\cite{bertizzolo2020arena}, which can all be used to generate massive datasets under a variety of network configurations and \gls{rf} conditions.

\noindent
$\bullet$ We provide the first demonstration of an O-RAN data-driven control loop in a \textit{large-scale experimental testbed using open-source, programmable} \gls{ran} and \gls{ric} components.
%Precisely,
We deploy O-RAN on the Colosseum network emulator and use it to control multiple network slices instantiated on 4~\gls{sdr} \glspl{bs} serving 40~\gls{sdr} \glspl{ue}.

\noindent
$\bullet$ We develop a set of \gls{drl} agents as \gls{ric} xApps to optimize key performance metrics for different network slices through \textit{data-driven closed-control loops}.
Experimental results show that our \gls{drl} approach outperforms other control strategies improving 
spectral efficiency by up to~20\% and reducing buffer occupancy by up to~37\%.
We released the \gls{drl} agents and the $7\:\mathrm{GB}$ dataset used to train them.\footnote{\url{https://github.com/wineslab/colosseum-oran-commag-dataset}}
%The code of the \gls{drl} agents and the $7\:\mathrm{GB}$ dataset used to train them is available to the research community.\footnote{\url{https://github.com/wineslab/colosseum-oran-commag-dataset}}

\begin{figure*}
\setlength\belowcaptionskip{-0.5cm}
	\centering
	\includegraphics[width=\textwidth]{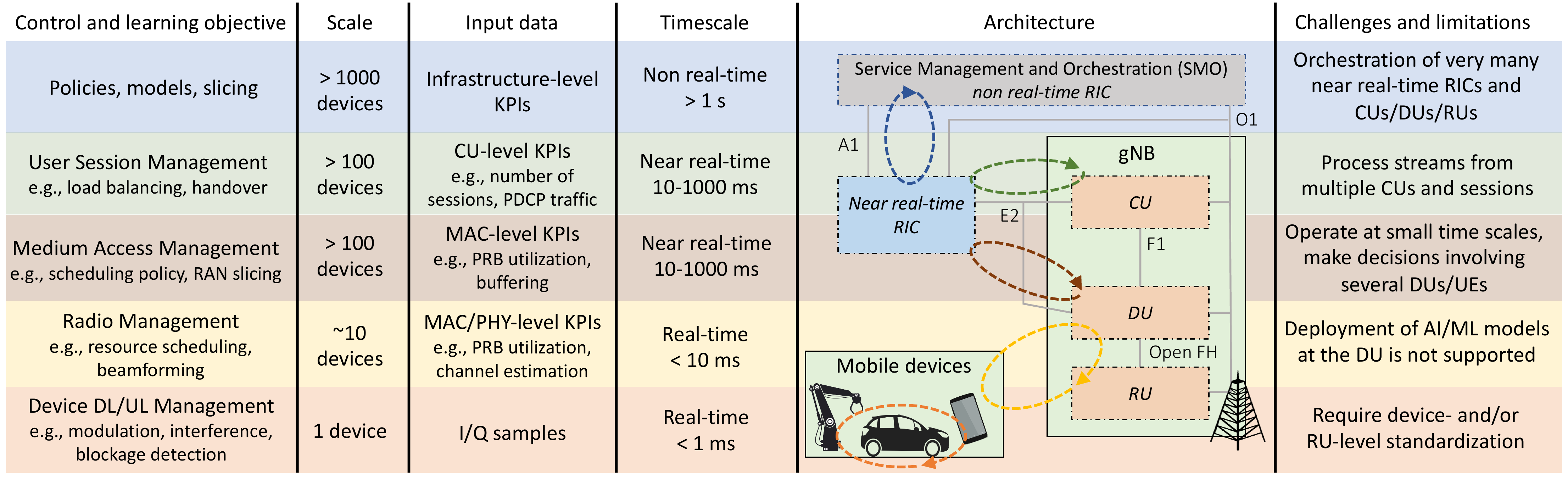}
	\caption{Learning-based closed-control loops in an O-RAN architecture.}
	\label{fig:loops}
\end{figure*}

The remainder of this article is organized as follows. 
We first discuss how intelligent control schemes can be embedded in the O-RAN architecture.
We then present how experimental testbeds can foster the development of data-driven solutions.
Finally, we present our experimental evaluation and draw our conclusions.

\vspace{-0.3cm}
\section{Intelligent Wireless Architectures}
\label{sec:intelligent}

Openness, programmability, and disaggregation are key enablers of data-driven applications. However, they are only the first step toward the seamless integration of \gls{ai} and \gls{ml}-based control loops in cellular networks.
Typically, data-driven approaches involve several
%well-established
steps, ranging from data collection and processing, to training, model deployment and closed-loop control and testing. 

This section illustrates how O-RAN is steering~5G deployments to bring intelligence to the network, by defining a practical architecture for the swift execution of data-driven operations, and discusses extensions to control procedures not currently considered by O-RAN.

% introduces possible extensions to implement the above pipeline in control loops that O-RAN is not currently addressing.

\textbf{Data Handling and Training Procedures}. The effectiveness of data-driven approaches heavily depends on how data is handled, starting from data collection and aggregation at the \gls{ran} (where data is generated) to the point where it is processed for model training and inference. However, collecting and moving large amounts of data might result in significant overhead and latency costs. Hence, data-driven architectures must cope with tradeoffs between centralized approaches---providing a comprehensive view of the state of the network at the cost of overhead and latency---and distributed ones---which operate at the edge only, gather data from a small number of sources while enjoying low latency~\cite{polese2018machine}.

In this context, the O-RAN \gls{ml} specifications introduce standardized interfaces (e.g., O1) to collect and distribute data across the entire infrastructure as well as operational guidelines for the deployment of \gls{ml} and \gls{ai} solutions in the network~\cite{oran-ml}. These include practical considerations on how, where and when models can be trained, tested and eventually deployed in the network. First, \gls{ai}/\gls{ml} models are made available to operators via a marketplace system similar to that of the well-established \Gls{nfv} \Gls{mano} architecture, where models are stored in a catalog together with details on their control objectives, required resources, and expected inputs and outputs. Second, data-driven solutions must be trained and validated offline to avoid causing inefficiencies---or even outages---to the \gls{ran}.
Indeed, since \gls{ai}/\gls{ml} techniques usually rely upon a randomized initialization, O-RAN requires all \gls{ml} models to be trained and validated offline before their deployment~\cite{oran-ml}.
%, and allows online learning only if the model is loaded with pre-trained weights already.
As we will discuss next, albeit shielding the network from unwanted behavior, this requirement also limits the effectiveness of such approaches, especially the online ones.
Online \gls{ai}/\gls{ml} techniques could still be used in O-RAN compliant architectures by allowing models to be trained with offline data in the non real-time RIC, and then perform online learning in the near real-time \gls{ric}. The smaller time-scale of the control loops of the latter would in fact allow the online training pipeline to be fed with data collected in real time.

\textbf{Control Loops}. Figure~\ref{fig:loops} portraits
%provides an overview of
how intelligence can be embedded at different layers and entities of a disaggregated cellular network together with the challenges and limitations of doing so. Each closed-control loop optimizes \gls{ran} parameters and operations by running at different timescales, with different number of \glspl{ue}, and using different sources for the input data.
%In this context,
The O-RAN Alliance is also looking into how to standardize the data-driven workflows for these control loops. As of this writing, O-RAN only considers non and near real-time loops, while real-time loops are left for future studies. 
Figure~\ref{fig:loops} also depicts the additional inference timescale below $1\:\mathrm{ms}$ to process raw I/Q samples and perform \gls{ai}-driven \gls{phy} layer tasks, currently not part of 
%the O-RAN architecture
O-RAN as it would require device- and/or \gls{ru}-level standardization. 

To better highlight the potential and limitations of the approach proposed by O-RAN, in the following we analyze each control loop individually, highlighting the role of each network component.
%as well as the data they have access to.
Finally, we discuss how the current O-RAN architecture can be extended to realize the control loops and applications illustrated in Fig.~\ref{fig:loops}.

\vspace{-.3cm}
\subsection{Non Real-time Control Loop}

The O-RAN Alliance defines \textit{non real-time} any control loop that operates on a \textit{timescale of at least one second.} As shown in Fig.~\ref{fig:loops}, this involves the coordination between the \textit{non real-time} and \textit{near real-time} \gls{ric}
%the non real-time \gls{ric} and the near real-time \gls{ric}
through the A1 interface. This control loop manages the orchestration of resources at the infrastructure level, making decisions and applying policies that impact thousands of devices. These actions can be performed using data-driven optimization algorithms processing data from multiple sources, and inference models deployed on the non real-time \gls{ric}~itself.

%An example of this is the non real-time \gls{ric}, which is in charge of deploying pre-trained models for inference in the near real-time \gls{ric}.

Practical examples of non real-time data-driven control include instantiating and orchestrating network slices, as well as selecting which pre-trained inference models in the catalog should be deployed to accomplish operator intents, and deciding in which near real-time \gls{ric} these models should be executed. Said decisions can be made according to a variety of factors, ranging from computational resources and data availability to minimum performance requirements to comply with \acrlongpl{sla}. Moreover, since the non real-time \gls{ric} is endowed with \acrlong{smo} capabilities, this control loop can also handle the association between the near real-time \gls{ric} and the \glspl{du}/\glspl{cu}. This is particularly useful in virtualized systems where \glspl{du} and \glspl{cu} are dynamically instantiated on-demand to match the requests and load of the \gls{ran}.
However, non real-time loops are challenging to actuate in practice because of the very many interactions among the non real-time \gls{ric} and the network elements, which require tight coordination, data collection and orchestration capabilities.

\vspace{-0.2cm}
\subsection{Near Real-time Control Loops}

\textit{Near real-time} control loops operate on a timescale \textit{between $10\:\mathrm{ms}$ and $1\:\mathrm{s}$}. As shown in Fig.~\ref{fig:loops}, they run between the near real-time \gls{ric} and two components of the \glspl{gnb}: The \gls{cu} and the \gls{du}. Because one near real-time \gls{ric} is associated to multiple \glspl{gnb}, these control loops can make decisions affecting up to thousands of \glspl{ue}, using user-session aggregated data and \gls{mac}/\gls{phy} layer \glspl{kpi}. \gls{ml}-based algorithms are implemented as external applications, i.e., \textit{xApps}, and are deployed on the near real-time \gls{ric} to deliver specific services such as inference, classification, and prediction pipelines to optimize the per-user quality of experience, controlling load balancing and handover processes, or the scheduling and beamforming design.
Challenges of near real-time control loops include the need to promptly make decisions in a matter of tens or hundreds of milliseconds for each of the several \glspl{cu} and \glspl{du} controlled by the \gls{ric}.

\vspace{-0.1cm}
\subsection{Real-time Control Loops}

A crucial component of the operations of a cellular network involves actions at a \textit{sub-$10\:\mathrm{ms}$---or even sub-ms---timescale.} In O-RAN, these operations are labeled as \textit{real-time control loops,} and mainly concern interactions between elements in the \gls{du}. Control loops at a similar timescale could also be envisioned to operate between the \gls{du} and the \gls{ru}, or at the \glspl{ue}.
%(although these cases are not natively covered by O-RAN).
However, as deploying \gls{ml} solutions at the \gls{du} is not currently supported, these loops are left for future extensions of the O-RAN specifications.

Finally, data-driven approaches at the lower layers of the protocol stack or at the device, i.e., involving sub-ms timescales, are extremely powerful and can be used for data-driven scheduling decisions~\cite{chinchali2018cellular} and for feedback-less detection of \gls{phy} layer parameters (e.g., modulation and coding scheme, and interference recognition)~\cite{oshea2017introduction}.
Overall, the fact that device-/\gls{ru}-level standardization is required for sub-$ms$ loops makes it very challenging to realize them in practice, thus limiting their applicability.
%However, sub-ms operations are out of the scope of the O-RAN architecture, as they require device- and/or \glspl{ru}-level standardization.

\begin{figure*}
\setlength\belowcaptionskip{-0.7cm}
	\centering
	\includegraphics[width=.95\textwidth]{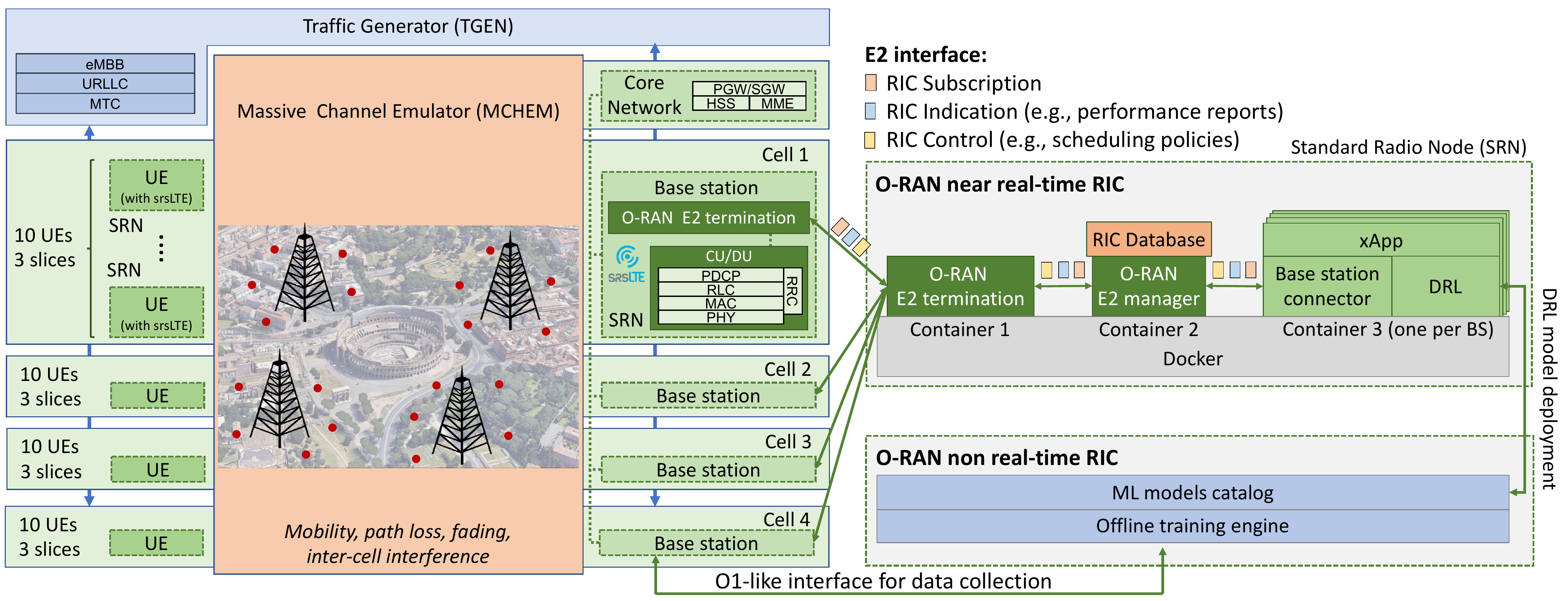}
	\caption{O-RAN integration in Colosseum.}
	\label{fig:colosseum}
\end{figure*}

\vspace{-0.2cm}
\section{Open Wireless Data Factories}
\label{sec:dataset}

Data-driven approaches aim at autonomously managing the network requiring little to no human intervention.
Training and testing algorithms and data-driven closed-control loop policies require large amounts of data gathered in diverse scenarios, with varying traffic patterns, requirements and user behaviors, so that the resulting policy is
%truly general and
effective when deployed in real networks.
%
%the O-RAN \gls{ml} workflow requires offline-trained and -tested models, as deploying online-only models could potentially introduce disruptive behaviors in the \gls{ran} in case they do not behave as expected.

Access to the massive amounts of data needed for training, however, is usually a privilege that only telecom operators enjoy. Owing to privacy and competition concerns, operators seldom share such data openly with the research community. As a consequence, researchers and practitioners are often constrained to rely on datasets collected in small laboratory setups, which seldom capture the variety and scale of real cellular deployments.
In the context of intelligent networking for NextG cellular systems, large-scale wireless testbeds are needed for developing, training and testing new data-driven solutions, serving as \textit{open wireless data factories} for the community.
%Indeed, these open platforms can facilitate massive data collection in realistic and diverse wireless deployments, and are also generally available to the research community at large~\cite{bonati2020open}.
Such open platforms would facilitate massive data collection in realistic and diverse wireless deployments~\cite{bonati2020open}.

The city-scale platforms of the U.S.\ National Science Foundation \acrshort{pawr} program promise to be a valuable tool to provide the community with the desired diversity of scenarios and scale.
The program is currently supporting three open testbeds representative of a variety of wireless use-cases, ranging from state-of-the-art \glspl{sdr} and massive \gls{mimo} communications (POWDER, in Salt Lake City, UT~\cite{breen2020powder}), to ultra-high capacity
%(e.g., mmWave)
and low-latency wireless networks (COSMOS, in New York City~\cite{raychaudhuri2020cosmos}), and to aerial wireless communications (AERPAW, in the Research Triangle of North Carolina~\cite{sichitiu2020aerpaw}).
All three platforms provide users with
%wireless
data generation and analysis tools~\cite{bonati2020open}.
Arena is a \gls{sdr} ceiling testbed that allows to study \gls{mimo}, cellular and \gls{iot} applications with up to 64 antennas deployed in an $8\times8$ grid in an office space~\cite{bertizzolo2020arena}.

Another instrument for wireless research at scale
%that has been recently made available to the research community at large
is Colosseum, the world's largest wireless network emulator with hardware-in-the-loop.
%, developed by DARPA and now available to the wireless research community through an NSF-funded effort at Northeastern University.
Colosseum includes 128 compute nodes, called \glspl{srn}, equipped with \acrshort{usrp}~X310 \glspl{sdr} that can be used to run generic protocol stacks. These are connected in a mesh topology through~128 additional \acrshortpl{usrp}~X310 of the \gls{mchem} for emulating realistic \gls{rf} scenarios.
% , and interfering devices. 
The wireless channel between each pair of devices is modeled through complex-valued \acrlong{fir} filter taps. In this way, scenarios are able to capture effects such as path loss, multi-path and fading as if the \glspl{sdr} were operating in a real \gls{rf} environment.
Colosseum is also equipped with an edge datacenter, with $900\:\mathrm{TB}$ of storage and the capability of processing \gls{rf} data at a rate of $52\:\mathrm{TB/s}$, enabling massive data collection and testing of \gls{ml} algorithms on heterogeneous networks.
%and devices.

%\vspace{-0.2cm}
\section{Use Case: Scheduling Control in Sliced 5G Networks through the O-RAN RIC}
\label{sec:usecase}

This section showcases an example of a data-driven closed-loop control implemented using the O-RAN Software Community near real-time \gls{ric} and an open cellular stack on Colosseum (Fig.~\ref{fig:colosseum}).
We demonstrate the feasibility of a closed-control loop where \gls{drl} agents running in xApps on the near real-time \gls{ric} select the best-performing scheduling policy for each \gls{ran} slice.

%We integrated several components of the open source code available through the O-RAN Software Community, a joint effort of the O-RAN Alliance and of the Linux Foundation to provide open source implementations of the O-RAN infrastructure.

\textbf{Experimental Scenario}. We have emulated a 5G network with 4 \glspl{bs} and 40 \glspl{ue} (Fig.~\ref{fig:colosseum}, left) in the dense urban scenario of Rome, Italy.
The locations of the \glspl{bs} have been extracted from OpenCelliD (a database of real-world cellular deployments) and cover an area of $0.11\:\mathrm{km^2}$.
Downlink and uplink frequencies have been set to $0.98$ and $1.02\:\mathrm{GHz}$, respectively; the channel bandwidth to $3\:\mathrm{MHz}$.
While these parameters might be atypical for 5G, their choice depends on the Colosseum environment. We note, however, that this does not affect our findings on how data-driven solutions improve the \gls{ran} performance.

We consider a multi-slice scenario in which \glspl{ue} are statically assigned to a slice of the network and request three different traffic types, i.e., high capacity \gls{embb}, \gls{urllc}, and \gls{mtc}.
This reflects the case, for instance, of telecom operators providing different levels of service to different devices (e.g., \gls{mtc} service to \gls{iot}-enabled devices, or \gls{urllc} to devices for time-critical applications).
The \glspl{bs} serve each slice with a dedicated---and possibly different---scheduling policy, selecting among Proportionally Fair (PF), Waterfilling (WF), and Round Robin (RR)~\cite{bonati2020cellos}. 
We also consider the case where
%\gls{ran} slicing is in place and
the number of \glspl{prb} allocated to each slice varies over time~\cite{doroInfocom2019Slicing, doro2020sledge}.

We used srsLTE to implement our softwarized cellular network. 
This open-source framework, which has recently been renamed ``srsRAN'' to reflect a new focus toward~5G NR, provides a full-stack implementation of \glspl{bs} and \glspl{ue}, as well as a lightweight core network.
Although this framework is not yet fully compliant with the NR specifications, we are confident that our \gls{drl}-based approach enabled by O-RAN can be easily extended to future NR-compliant versions of this (or of any other) software where \glspl{bs} expose control interfaces to the network.
%
%co-located with the \gls{bs}.
For ease of prototyping, we co-located the core network on the same \gls{srn} that also runs the \gls{bs} application. For the purposes of our work, this setup is equivalent to deploying the core network on a dedicated \gls{srn} (see Fig.~\ref{fig:colosseum}).
We extended the \gls{bs} implementation to include network slicing capabilities and additional scheduling policies~\cite{bonati2020cellos}. The scenario we considered concerns pedestrian user mobility 
%resulting in \gls{mchem} producing 
with time-varying path-loss and channel conditions. Traffic among \glspl{bs} and \glspl{ue} is generated through the Colosseum \gls{tgen}, configured to send different traffic types to \glspl{ue} of different slices, i.e., \gls{embb} ($1\:\mathrm{Mbps}$ constant bitrate traffic), \gls{urllc} (Poisson traffic, with $10\:\mathrm{pkt/s}$ of $125\:\mathrm{bytes}$) and \gls{mtc} (Poisson traffic, with $30\:\mathrm{pkt/s}$ of $125\:\mathrm{bytes}$). For each \gls{bs}, the \gls{ue}-slice allocation is as follows: \gls{embb} and \gls{urllc} slices serve 3 \glspl{ue} each, while \gls{mtc} slices serve 4~\glspl{ue}.
We embedded the \gls{drl} agents into xApps running in the near real-time \gls{ric} (right of Fig.~\ref{fig:colosseum}), for a total of 12 \gls{drl} agents running in parallel
%(controlling 4 \glspl{bs} with 3 slices each)
and making decisions with a time granularity of~$500\:\mathrm{ms}$.
%Agents make scheduling decisions for the network slices with a time granularity of $500\:\mathrm{ms}$.
%
Agents connect with the network \glspl{bs} through the O-RAN E2 interface. This interface is composed of two elements: The \acrlong{ap}, and the \gls{sm}~\cite{oran-e2}. The former defines the set of messages that the near real-time \gls{ric} and the \gls{ran} nodes can exchange, and the procedures for the \gls{ran} node subscription to the \gls{ric}. The \gls{sm}, instead, defines which parameters of the \gls{ran} nodes can be controlled by the \gls{ric} to achieve a given closed-loop control objective. Specifically, the E2 interface exposes analytics and the scheduler policy selection using a custom \gls{sm}.
As shown in Fig.~\ref{fig:colosseum}, xApps interface with the \glspl{bs}
%they control
through the O-RAN E2 manager, which ultimately connects with the \glspl{bs} via the E2 interface. Other components of the \gls{ric} include the \gls{ric} database, which keeps entries on the connected \glspl{bs}, the training engine, and the \gls{ml} model catalog, which deploys the \gls{drl} model chosen by the telecom operator on the near real-time \gls{ric}.
Finally, messages internal to the \gls{ric} are managed by the \acrlong{rmr}, a library which associates message types to destination endpoints.

\textbf{DRL Agent Training}. To train our \gls{drl} agents we generated some $7\:\mathrm{GB}$ of training data of various performance metrics (e.g., throughput and bit error rate), system state information (e.g., transmission queue size, signal-to-interference-plus-noise ratio, and channel quality information) and resource allocation strategies (e.g., slicing and scheduling policies) by running a total of 89 hours of experiments on Colosseum. Each \gls{drl} agent has been trained via the \gls{ppo} algorithm to manage a single slice for fine-grained and flexible control of the whole cellular network. 
Agents have been trained under network configurations obtained by varying the distance between \glspl{bs} and \glspl{ue} and the mobility of the \glspl{ue}. Testing has been performed in the most challenging setup, which includes the random mobility of the \glspl{ue}. 
Although the training is performed with the same topology configuration, we notice that our agents are topology-independent, as each of them controls a single slice for a given \gls{bs}.
Specifically, agents process the performance metrics received by the \gls{bs} they are controlling---which possibly expresses the performance of several \glspl{ue}---through an encoder. This allows them to cast the dimensionality of the input data to a fixed size and to process it regardless of the number of active \glspl{ue} of the slice.
As a consequence, the \gls{drl} agents do not need to be aware of the number of \glspl{ue} and \glspl{bs} in the network, which makes our approach general and scalable.
%(Due to space limitations, we refrain from discussing details of this actor-critic \gls{drl} framework in this article.)
Through the \gls{ric} Indication messages sent via the O-RAN E2 interface (Fig.~\ref{fig:colosseum}), the agent is fed real-time performance measurements of the slice it controls. These messages generate an overhead of $72\:\mathrm{bytes/s}$ per \gls{ue}.
%in our prototype.
Data goes through an encoder for dimensionality reduction and
%
%The output
is then used by the agent to identify the state of the system.
The agent uses a fully connected neural network with 5~layers and 30~neurons each to determine the best scheduling policy for the corresponding slice.
This policy is then signaled to the corresponding \gls{bs} through \gls{ric} Control messages sent via the E2 interface.
%(Fig.~\ref{fig:colosseum}).
The reward of the agents depends on the specific slice and the corresponding \gls{kpi} requirements. Specifically, \gls{embb} and \gls{mtc} agents have been trained to maximize the throughput of \glspl{ue}; the \gls{urllc} agent has been trained to minimize latency by allocating resources (i.e., \glspl{prb}) as quickly as possible.
%
%To fully
To comply with O-RAN directives, we have trained the \gls{drl} agents offline
in the non real-time \gls{ric}, which also performs the initial data-collection and deploys the model in the near real-time \gls{ric}.
We have then tested them on the emulated Colosseum scenario. 

\begin{figure}[t]
\setlength\belowcaptionskip{-1cm}
	\centering
	\includegraphics[width=\columnwidth]{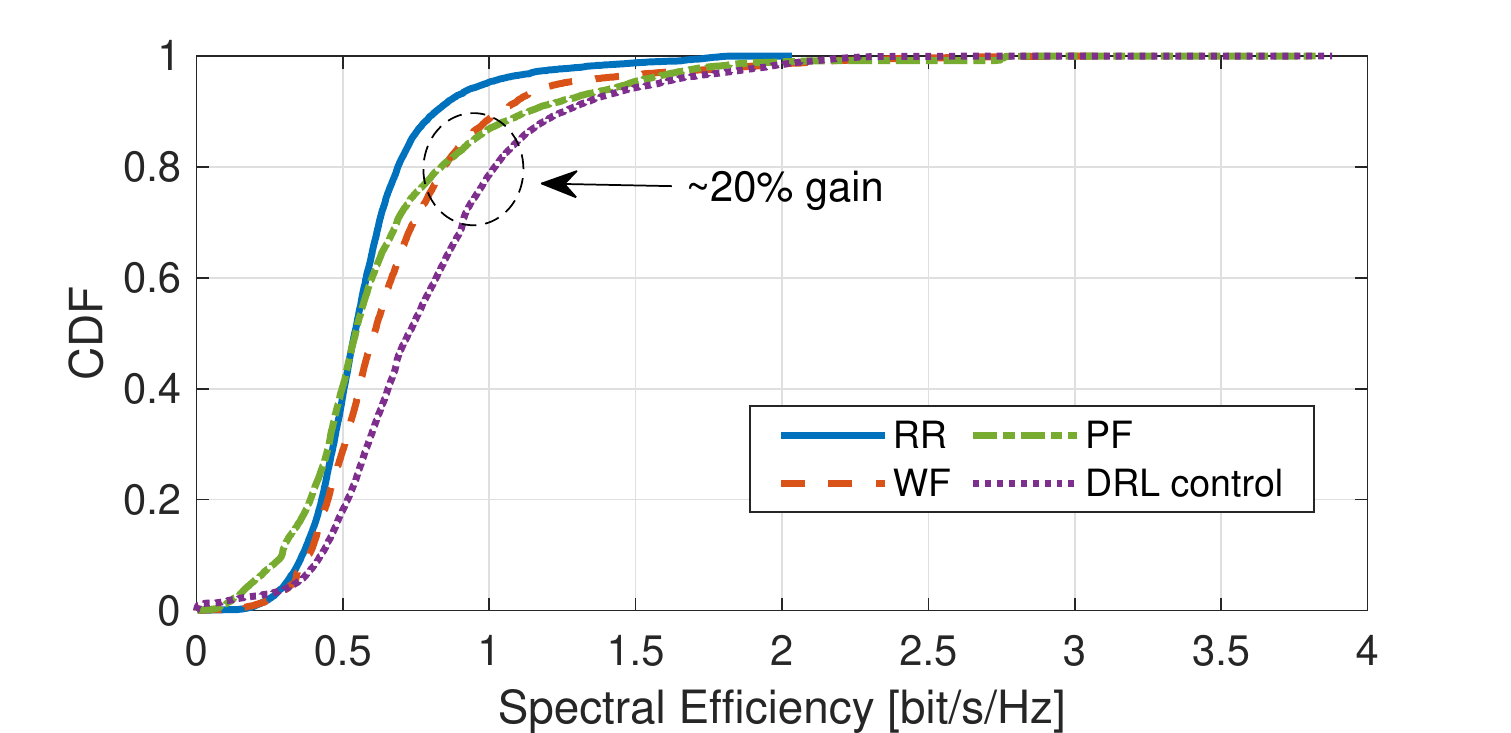}
	\caption{Downlink spectral efficiency of the \acrshort{embb} slice for different scheduling policies and with DRL control.}
	\label{fig:spectral}
\end{figure}

\textbf{Experimental Results}. 
%To demonstrate the benefits of closed-loop data-driven optimization in O-RAN applications, in 
Figure~\ref{fig:spectral} shows the \gls{cdf} of the downlink spectral efficiency of the \gls{embb} slice.
\vspace{-0.1cm}
We compare the performance of the network when \gls{drl} agents dynamically select the best scheduling strategy among RR, PF and WF against the case where scheduling strategies are fixed over time. Our results clearly indicate that data-driven optimization outperforms fixed policies by delivering gains in spectral efficiency that are up to 20\% higher than that of the best performing static policy. This is due to the fact that \gls{embb} traffic requires high data-rates and \gls{drl} agents are capable of dynamically adapting scheduling decisions to the current network state and traffic demand.

\begin{figure}[ht]
\setlength\belowcaptionskip{-0.2cm}
	\centering
	\includegraphics[width=0.85\columnwidth]{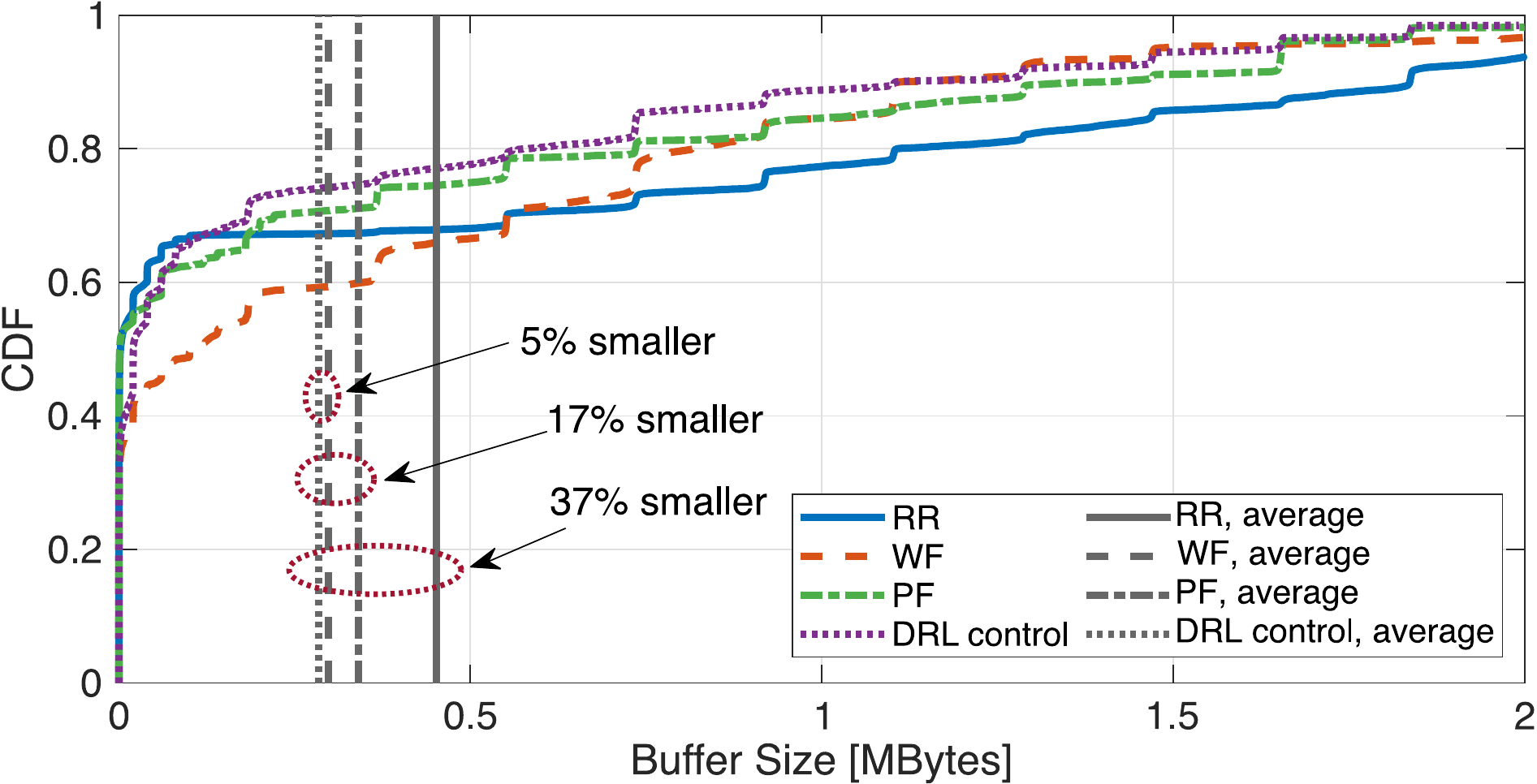}
	\caption{Downlink buffer size of the URLLC slice for different scheduling policies and with DRL control.}
	\label{fig:buffer}
\end{figure}

Figure~\ref{fig:buffer} shows the \gls{cdf}
%and average values
of the downlink buffer size for the \gls{urllc} slice under different scheduling policies.
Low buffer size indicates timely data delivery to requesting \glspl{ue}; higher buffer size results in higher latency due to packets waiting in the queue.
Results show that \gls{drl} agents serve the \glspl{ue} faster than the static policies, resulting in a lower latency. 
Particularly, the average buffer size of the \gls{urllc} slice when using \gls{drl} control is $37\%$, $5\%$ and $17\%$ smaller than that of the RR, WF and PF scheduling policies, respectively. The \gls{drl} agent also significantly outperforms the WF policy between the $50$th and $90$th percentiles.

\begin{figure}[t]
\setlength\belowcaptionskip{-0.7cm}
	\centering
	\includegraphics[width=0.9\columnwidth]{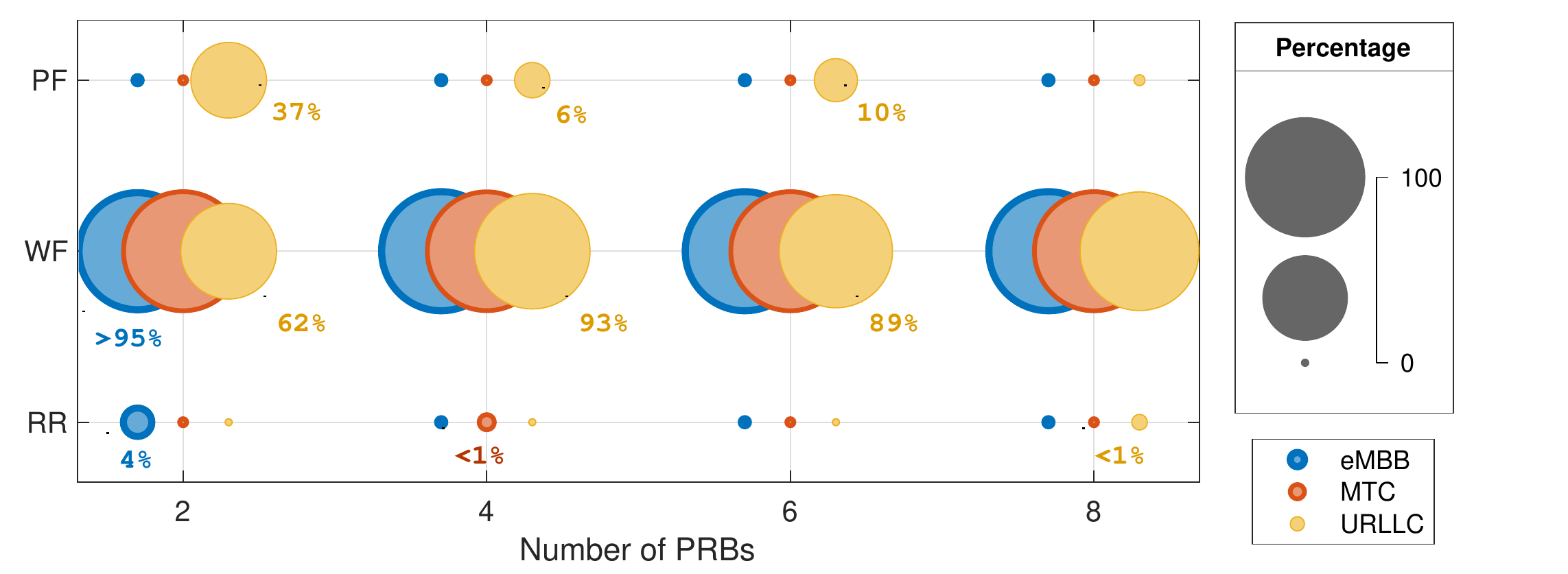}
	\caption{DRL action selection distribution vs.\ number of slice PRBs. Values \mbox{$>99\%$} (big circles) or \mbox{$<0.5\%$} (small circles) are omitted.}
	\label{fig:actions}
\end{figure}

Figure~\ref{fig:actions} depicts how often \gls{drl} agents select specific scheduling policies as a function of the number of \glspl{prb} of each slice. The bigger the circle, the higher the probability of selecting a given policy. We observe that \gls{mtc} and \gls{embb} \gls{drl} agents select WF with $99\%$ probability.
Also, \gls{embb} agents select RR with $4\%$ probability when only a few \glspl{prb} are allocated to the slice.
On the contrary, \gls{urllc} \gls{drl} agents are likely to select both PF and WF scheduling policies even when more \glspl{prb} are available. These results show that adapting control strategies to current network state and traffic requirements is essential to achieve remarkable performance improvements (Fig.~\ref{fig:spectral} and~\ref{fig:buffer}).
\gls{drl} agents dynamically select the best performing scheduling strategy based on available resources and current network state, providing performance gains simply unattainable with static scheduling policies.

\vspace{-0.2cm}
\section{Conclusions}
\label{sec:conclusions}

In this article we provide a path and a demonstration of the feasibility of integrating closed-control loops in cellular networks. 
%effectively implementing the vision of self-optimizing autonomous networks. 
We first review key enablers, namely, virtualization, disaggregation, openness and reprogrammability of NextG cellular networks, using O-RAN as an exemplary technology. We then discuss which data-driven control loops can be implemented, their timescale, and whether the current O-RAN architecture supports them. We finally show how large-scale experimental testbeds can be used to develop and validate data-driven algorithms by deploying a DRL-based O-RAN \gls{ric} on Colosseum. Our results show that using closed-control loops can provide a strong foundation toward the full realization of future generation, data-driven, autonomous, and self-optimizing cellular networks.

%\vspace{-0.2cm}
\balance
\bibliographystyle{IEEEtran}
\bibliography{biblio.bib}

\vspace{-1.3cm}
\begin{IEEEbiographynophoto}{Leonardo Bonati} [S'19] is a Ph.D.\ candidate at Northeastern University. He received his M.S.\ in Telecommunication Engineering from the University of Padova, Italy in 2016. His research focuses on softwarized NextG systems.
\end{IEEEbiographynophoto}

\vspace{-1.3cm}
\begin{IEEEbiographynophoto}
{Salvatore D'Oro} [M'17] is a Research Assistant Professor with the Institute for the Wireless Internet of Things at Northeastern University, USA. He received his Ph.D.\ from the University of Catania in 2015. He serves on the technical program committee of IEEE INFOCOM. His research focuses on optimization and learning for NextG systems.
\end{IEEEbiographynophoto}

\vspace{-1.3cm}
\begin{IEEEbiographynophoto}{Michele Polese}
[M'20] is a research scientist at Northeastern University. He obtained his Ph.D.\ from the University of Padova, Italy, in 2020, where he also was a postdoctoral researcher and adjunct professor. His research focuses on architectures for wireless networks.
\end{IEEEbiographynophoto}

\vspace{-1.3cm}
\begin{IEEEbiographynophoto}{Stefano Basagni}
[SM'06] received a Ph.D.\ in electrical engineering from the University of Texas at Dallas in 2001
and a Ph.D.\ in computer science from the University of Milano, Italy in 1998.
He is with the Institute for the Wireless Internet of Things and a professor in the ECE Department at Northeastern University. His research concerns mobile networks and wireless communications systems.
\end{IEEEbiographynophoto}

\vspace{-1.3cm}
\begin{IEEEbiographynophoto}{Tommaso Melodia}
[F’18] received a Ph.D.\ in Electrical and Computer Engineering from the Georgia Institute of Technology in 2007. He is the William Lincoln Smith Professor at Northeastern University, the Director of the Institute for the Wireless Internet of Things, and the Director of Research for the PAWR Project Office. His research focuses on wireless networked systems.
\end{IEEEbiographynophoto}

\end{document}